\def\bds#1{\boldsymbol{#1}}

\def\rfr#1{(\ref{#1})}
\def\rfrs#1#2{(\ref{#1})-(\ref{#2})}

\def\dert#1#2{\frac{{{d}}{#1}}{{{d}}{#2}}}              % derivate parziali e totali prima e seconda

\def\bar{\begin{eqnarray}}
\def\ear{\end{eqnarray}}

\def\eqi{\begin{equation}}
\def\eqf{\end{equation}}
\def\eqia{\begin{eqnarray}}
\def\eqfa{\end{eqnarray}}
\def\rp#1#2{{#1\over#2}}
\def\ct#1{\cite{#1}}
\def\lb#1{\label{#1}}

\documentclass[11pt]{article}
\usepackage{amsmath,amsthm,amscd,amssymb}
\usepackage{latexsym}
\usepackage{graphicx,epsfig}

\begin{document}

\noindent{\bf \LARGE{The post-Newtonian mean anomaly advance as
further post-Keplerian parameter in pulsar binary systems.}}
\\
\\
\\
{Lorenzo Iorio}, {\it FRAS, DDG}\\
{\it Viale Unit${\rm \grave{a}}$ di Italia 68, 70125\\Bari, Italy
\\e-mail: lorenzo.iorio@libero.it}

\begin{abstract}
The post-Newtonian gravitoelectric secular rate of the mean
anomaly ${\mathcal{M}}$ is worked out for a two-body system in the
framework of the General Theory of Relativity. The possibility of
using such an effect, which is different from the well known
decrease of the orbital period due to gravitational wave emission,
as a further post-Keplerian parameter in binary systems including
one pulsar is examined. The resulting effect is almost three times
larger than the periastron advance $\dot\omega$. E.g., for the
recently discovered double pulsar system PSR J0737-3039 A+B it
would amount to -47.79 deg yr$^{-1}$. This implies that it could
be extracted from the linear part of a quadratic fit of the
orbital phase because the uncertainties both in the linear drift
due to the mean motion and in the quadratic shift due to the
gravitational wave are smaller. The availability of such
additional post-Keplerian parameter would be helpful in further
constraining the General Theory of Relativity, especially for such
systems in which some of the other post-Keplerian parameters can
be measured with limited accuracy. Moreover, also certain
pulsar-white dwarf binary systems, characterized by circular
orbits like PSR B1855+09 and a limited number of measured
post-Keplerian parameters, could be used for constraining
competing theories of gravity.
\end{abstract}

\section{The post-Newtonian rate of the mean anomaly}
According to the Einsteinian General Theory of Relativity (GTR),
the post-Newtonian gravitoelectric two-body acceleration of order
$\mathcal{O}(c^{-2})$ (1PN) is, in the post-Newtonian centre of
mass frame (see \ct{Dam85} and, e.g., \ct{port04} and references therein)\eqi
{\bds a}_{\rm GE}=\rp{Gm}{c^2
r^3}\left\{\left[\rp{Gm}{r}(4+2\nu)-(1+3\nu)v^2+\rp{3\nu}{2r^2}(\bds
r\cdot \bds v )^2\right]\bds r+(\bds r\cdot\bds v)(4-2\nu)\bds v
\right\},\lb{acc}\eqf where $\bds r$ and $\bds v$ are the relative
position and velocity vectors, respectively, $G$ is the Newtonian
constant of gravitation, $c$ is the speed of light, $m_1$ and $m_2$
are the rest masses of the two bodies, $m\equiv m_1+m_2$ and $\nu\equiv m_1
m_2/m^2<1.$

The orbital phase can be characterized by the mean anomaly
$\mathcal{M}$ defined as
\eqi\mathcal{M}=n(t-T_0)\lb{meananom},\eqf where the
unperturbed mean motion $n$ is defined as \eqi
n=\rp{2\pi}{P_b},\eqf $P_b$ is the anomalistic period, i.e. the
time elapsed between two consecutive pericentre crossings, which
is $2\pi\sqrt{a^3/Gm}$ for an unperturbed Keplerian ellipse, and
$T_0$ is the date of a chosen pericentre passage.

The variation of the mean anomaly can be written, in general, as
\eqi \dert{\mathcal{M}}{t}=n-2\pi\left(\rp{\dot
P_b}{P_b^2}\right)(t-T_0)-\rp{2\pi}{P_b}\left(\dert{T_0}{t}\right).\lb{general}\eqf
 The second term of the right-hand side of
\rfr{general} accounts for any possible variation of the
anomalistic period. The third term, induced by any small
perturbing acceleration with respect to the Newtonian monopole,
whether relativistic or not, is the
change of the time of the pericentre passage, which we will define as
\eqi\dert{\xi}{t}\equiv -\rp{2\pi}{P_b}\left(\dert{T_0}{t}\right).\eqf It can be
calculated with the aid of the Gauss\footnote{For a different approach based on a modified form of the Lagrange planetary equations see \ct{Cal97}.}  perturbative equation
\ct{roy88} \eqi\dert{\xi}{t}=
-\rp{2}{na}R\left(\rp{r}{a}\right)-\sqrt{1-e^2}\left(\dert{\omega}{t}+\cos
i\dert{\Omega}{t}\right),\lb{manom} \eqf where $R$ is the radial
component of the disturbing acceleration, $a,e,i,\Omega$ and
$\omega$ are the semimajor axis, the eccentricity, the
inclination, the longitude of the ascending node and the argument
of pericentre, respectively, of the orbit. In order to obtain the
secular effects, we must evaluate  the right-hand-side of
\rfr{manom} on the unperturbed Keplerian ellipse and, then,
average the result over one orbital revolution. In regard to the
physical interpretation of $d\xi/dt$, the apsidal line can be
crossed at a different time with respect to the unperturbed
Keplerian case either because there is an additional radial
acceleration which accelerates/decelerates the moving particle or
because the pericentre is no more fixed in space. The first case
is accounted for by the first term of the right-hand side of
\rfr{manom}, while the second term of the right-hand side of
\rfr{manom} explains the second case. Indeed, for, e.g., an inward
radial acceleration which increases the gravitational strength the
particle moves more rapidly and the pericentre, assumed fixed in
space, is reached in advance with respect to the unperturbed case:
$dT_0/dt<0$. If the angle $\omega+\Omega\cos i$, which defines an angular variable around the axis of the orbital angular momentum, increases, the passage at pericentre
will occur later with respect to the unperturbed case: $d
T_0/dt>0$.

%\footnote{As we
%will see later, $P_{\rm b}$ is one of the orbital parameters of
%the binary pulsar systems which are better determined from the
%observations in a purely phenomenological way, independently of
%any gravitation theory.
%
% in general, it can also
%include post-Newtonian corrections as that calculated in
%\ct{AHIP}: see also (10), (12) and (14) of \ct{Calura}.

We will now consider \rfr{acc} as perturbing acceleration.
 Let us start with the first term of the right-hand-side of
\rfr{manom}. By defining \eqi\left\{
\begin{array}{lll}\lb{defz}
A&\equiv &\rp{(Gm)^2}{c^2}(4+2\nu),\\\\
B&\equiv &-\rp{Gm}{c^2}(1+3\nu),\\\\
C&\equiv &\rp{Gm}{c^2}\left(4-\rp{\nu}{2}\right),
\end{array}
\right. \eqf it is possible to obtain from \rfr{acc}\eqi R_{\rm
GE}=\rp{A}{r^3}+B\left(\rp{v^2}{r^2}\right)+C\left(\rp{\dot r}{r^2}\right).\lb{radial}\eqf
Now the term $-2 R r/na^2$, with $R$ given by \rfr{radial}, must
be evaluated on the unperturbed Keplerian ellipse characterized by
\eqi\left\{
\begin{array}{lll}\lb{kepl}
r&=&\rp{a(1-e^2)}{1+e\cos f},\\\\
\dot r&=&\rp{nae\sin f}{\sqrt{1-e^2}},\\\\
v^2&=&\rp{n^2 a^2}{(1-e^2)}(1+e^2+2e\cos f)
\end{array}
\right. \eqf where $f$ is the true anomaly, and averaged over one
orbital period by means of
\begin{equation}
\rp{d t}{P_b}=\rp{r^2 df}{2\pi a^2\sqrt{1-e^2}}.
\end{equation}
Thus, \eqi -\left(\rp{2}{na}R_{\rm
GE}\rp{r}{a}\right)\left(\rp{dt}{P_b}\right)=-\rp{1}{na^4\pi\sqrt{1-e^2}}\left[A+Brv^2+Cr(\dot
r )^2\right]df .\lb{ert}\eqf In the expansion of $r$ in \rfr{ert}
the terms of order $\mathcal{O}(e^4)$ are retained. The final
result is \eqi\left\langle -\rp{2}{na}R_{\rm
GE}\rp{r}{a}\right\rangle_{P_b}= \rp{nGm}{c^2
a\sqrt{1-e^2}}H(e;\nu),\lb{inter}\eqf with \eqia H\simeq
&-&2(4+2\nu)+(1+3\nu)\left(2+e^2+\rp{e^4}{4}+\rp{e^6}{8}\right)-\nonumber\\
&-&\left(4-\rp{\nu}{2}\right)\left(e^2 +\rp{e^4}{4}+\rp{e^6}{8}
\right).\eqfa

The post-Newtonian gravitoelectric secular rate of pericentre is
independent of $\nu$ and is given by the well known formula
\eqi\left.\dert{\omega}{t}\right|_{\rm GE }=\rp{3nGm}{c^2
a(1-e^2)}\lb{perge},\eqf while there are no secular effects on the
node.

The final expression for the post-Newtonian secular rate of the
mean anomaly is obtained by combining \rfrs{inter}{perge} and by
considering that, for a two-body system, it is customarily to
write
\eqi\rp{nGm}{c^2}=\left(\rp{P_b}{2\pi}\right)^{-5/3}(T_{\odot}\mathbb{M})^{2/3},\eqf
where $\mathbb{M}=m/{\rm M}_{\odot}$ is the sum of the masses in
units of solar mass and $T_{\odot}=G{\rm
M}_{\odot}/c^3=4.925490947\times 10^{-6}$ s. It is
\eqi\left.\dert{\mathcal{\xi}}{t}\right|_{\rm
GE}=-9\left(\rp{P_b}{2\pi}\right)^{-5/3}(T_{\odot}\mathbb{M})^{2/3}(1-e^2)^{-1/2}F(e;\nu)\lb{fine}\eqf
with\eqi
F=\left[\left(1+\rp{e^2}{3}+\rp{e^4}{12}+\rp{e^6}{24}+...\right)-\rp{2}{9}\nu\left(
1+\rp{7}{4}e^2+\rp{7}{16}e^4+\rp{7}{32}e^6+...\right)\right].\lb{finee}\eqf
Note that \rfr{fine} is negative because \rfr{finee} is always
positive; thus the crossing of the apsidal line occurs at a later
time with respect to the Kepler-Newton case.
%The importance of this fact will become clearer
%in Section \ref{concl}.

Note that, for $\nu\rightarrow 0$, i.e. $m_1\ll m_2$, \rfr{fine}
does not vanish and, for small eccentricities, it becomes
\eqi\left.\dert{\mathcal{\xi}}{t}\right|_{\rm GE}\simeq
-\rp{9nGm_2}{c^2
a\sqrt{1-e^2}}\left(1+\rp{e^2}{3}\right),\lb{myles}\eqf which
could be used for planetary motion in the Solar System \ct{ior04}.
E.g., for Mercury it yields a secular effect of almost -130 arcsec
cy$^{-1}$. It is important to note that the validity of the
present calculations has also been numerically checked by
integrating over 200 years the Jet Propulsion Laboratory (JPL)
equations of motion of all the planets of the Solar System with
and without the gravitoelectric $1/c^2$ terms in the dynamical
force models \ct{esta} in order to single out just the
post-Newtonian gravitoelectric effects. They fully agree with
\rfr{myles} (E.M. Standish, private communication, 2004). Another
analytical calculation of the post-Newtonian general relativistic
gravitoelectric secular rate of the mean anomaly was performed
\ct{rubincam 1977} in the framework of the Lagrangian perturbative
scheme for a central body of mass $M$-test particle system. The
author of \ct{rubincam 1977} starts from the space-time line
element of the Schwarzschild metric written in terms of the
Schwarzschild radial coordinate $r^{'}$. Instead, \rfr{acc} and
the equations of motion adopted in the practical planetary data
reduction at, e.g. JPL, are written in terms of the standard
isotropic radial coordinate $r$ related to the Schwarzschild
coordinate by $r^{'}=r(1+GM/2c^2 r)^2$. As a consequence, the
obtained exact expression \eqi \left.\dert{\xi}{t}\right|_{\rm
GE }^{(\rm Rubincam )}=\rp{3nGM}{c^2a\sqrt{1-e^2}},\lb{rubi}\eqf contrary
to the pericentre case, agrees neither with \rfr{myles} nor with
the JPL numerical integrations yielding, e.g., a secular advance
of $+42$ arcsec cy$^{-1}$ for Mercury.  For a better understanding of such comparisons, let us note that both the numerical analysis by Standish and
of the author of   \ct{rubincam 1977} are based on the $\dot P_{\rm b}=0$ case; $n$ gets canceled by construction in the Standish calculation, while in \ct{rubincam 1977} the numbers are put just into \rfr{rubi}, which is the focus of that work.

\subsection{Testing gravitational theories with binary pulsars}
In general, in the pulsar's timing data reduction
process\footnote{For all general aspects of the binary pulsar
systems see \ct{wex01, sta03} and references therein.} five
Keplerian orbital parameters and a certain number of
post-Keplerian parameters are determined with great accuracy in a
phenomenological way, independently of any gravitational theory
\ct{wex01, sta03}.  The Keplerian parameters are the projected
semimajor axis $x=a\sin i/c$, where $i$ is the angle between the
plane of the sky, which is normal to the line of sight and is
assumed as reference plane, and the pulsar's orbital plane, the
eccentricity $e$, the orbital period $P_b$, the time of periastron
passage $T_0$ and the argument of periastron $\omega_0$ at the
reference time $T_0$. The most commonly used post-Keplerian
parameters are the periastron secular advance $\dot\omega$, the
combined time dilation and gravitational redshift due to the
pulsar's orbit $\gamma$, the variation of the anomalistic period
$\dot P_b$, the range $r$ and the shape $s$ of the Shapiro delay.
These post-Keplerian parameters are included in the timing models
\ct{wex01, sta03} of the so called Roemer, Einstein and Shapiro
$\Delta_{\rm R}, \Delta_{\rm E},\Delta_{\rm S}$
delays\footnote{For the complete expression of the timing models
including, e.g., also the delays occurring in the Solar System due
to the solar gravity see \ct{wex01, sta03}.} occurring in the
binary pulsar system\footnote{The aberration parameters $\delta_r$
and $\delta_{\theta}$ are not, in general, separately measurable.}
\eqi \left\{
\begin{array}{lll}\lb{delays}
\Delta_{\rm R}&=& x\sin\omega[\cos
E-e(1+\delta_r)]+x\cos\omega\sin
E\sqrt{1-e^2(1+\delta_{\theta})^2},\\\\
\Delta_{\rm E}&=&\gamma\sin E,\\\\
\Delta_{\rm S}&=&-2r\ln\{1-e\cos E-s[\sin\omega(\cos
E-e)+\sqrt{1-e^2}\cos\omega\sin E]\},
\end{array}
\right. \eqf where $E$ is the eccentric anomaly defined
as
%\footnote{Note that $e$ and $n$ here should be intended at 1PN
%level as (14) and (12) of \ct{Calura}.}
$E-e\sin E=\mathcal{M}$. $\cos E$ and $\sin E$ appearing in
\rfr{delays} can be expressed in terms of $\mathcal{M}$ by means
of the following elliptic expansions \ct{vinti} \eqi \left\{
\begin{array}{lll}\lb{dalam}
\cos E&=&-\rp{e}{2}+\sum_{j=1}^{\infty}\rp{2}{j^2}\rp{d}{de}[J_j (je)]\cos (j\mathcal{M}),\\\\\
\sin E&=&\rp{2}{e}\sum_{j=1}^{\infty}\rp{J_j(je)}{j}\sin
(j\mathcal{M}),
\end{array}
\right. \eqf where $J_j(y)$ are the Bessel functions defined as
\eqi \pi J_j(y)=\int_0^{\pi}\cos(j \theta-y\sin\theta)d\theta.\eqf
The relativistic secular advance of the mean anomaly \rfr{fine}
can be accounted for in the pulsar timing modelling by means of
\rfr{dalam}.
%This also would allow to include in the timing models
% $\dot{\mathcal{M}}_{\rm GE}$.

%apart from $P_b$, $\dot P_b$ and $\dot\omega$.

In a given theory of gravity, the post-Keplerian parameters can be
written in terms of the mass of the pulsar $m_p$ and of the
companion $m_c$. In general,  $m_p$ and $m_c$ are unknown; this
means that the measurement of only one post-Keplerian parameter,
say, the periastron advance, cannot be considered as a test of a
given theory of gravity because one would not have a theoretically
calculated value to be compared with the phenomenologically
measured one.  In GTR the previously quoted post-Keplerian
parameters are \ct{dam86} \eqi \left\{
\begin{array}{lll}
\dot\omega &=&
3\left(\rp{P_b}{2\pi}\right)^{-5/3}(T_{\odot}\mathbb{M})^{2/3}(1-e^2)^{-1},\\\\
\gamma &=&
e\left(\rp{P_b}{2\pi}\right)^{1/3}T_{\odot}^{2/3}\mathbb{M}^{-4/3}m_c(m_p+2m_c),\\\\
\dot P_b
&=&-\rp{192\pi}{5}T_{\odot}^{5/3}\left(\rp{P_b}{2\pi}\right)^{-5/3}\rp{
\left(1+\rp{73}{24}e^2+\rp{37}{96}e^4\right)}{(1-e^2)^{7/2}}\rp{m_p
m_c }{\mathbb{M}^{1/3}},\label{dpdt}\\\\
r&=&T_{\odot}m_c,\\\\
s&=&xT_{\odot}^{-1/3}\left(\rp{P_b}{2\pi}\right)^{-2/3}\rp{\mathbb{M}^{2/3}}{m_c}
\end{array}
\right. \eqf It is important to note that the relativistic
expression of $\dot P_b$ in \rfr{dpdt}, should not be confused
with $\left.\dot\xi\right|_{\rm GE}$ of  \rfr{fine}. Indeed, it
refers to the shrinking of the orbit due to gravitational wave
emission which vanishes in the limit $\nu\rightarrow 0$, contrary
to \rfr{fine} which expresses a different, independent phenomenon.
The measurement of two post-Keplerian orbital parameters allows to
determine $m_p$ and $m_c$, assumed the validity of a given theory
of gravity\footnote{This would still not be a test of the GTR
because the masses must be the same for all the theories of
gravity, of course.  }. Such values can, then, be inserted in the
analytical expressions of the remaining post-Keplerian parameters.
If the so obtained values are equal to the measured ones, or the
curves for the $2+N$, with $N\geq 1$, measured post-Keplerian
parameters in the $m_p-m_c$ plane all intersect in a well
determined $(m_p,m_c)$ point, the theory of gravity adopted is
consistent. So, in order to use the pulsar binary systems as
valuable tools for testing GTR the measurement of at least three
post-Keplerian parameters is required. The number of
post-Keplerian parameters which can effectively be determined
depends on the characteristics of the particular binary system
under consideration. For the pulsar-neutron star PSR B1913+16
system \ct{hul75} the three post-Keplerian parameters
$\dot\omega,\gamma$ and $\dot P_b$ were measured with great
accuracy. For the pulsar-neutron star PSR B1534+12 system
\ct{sta02} the post-Keplerian parameters reliably measured are
$\dot\omega,\gamma, r$ and $s$. For the recently discovered
pulsar-pulsar PSR J0737-3039 A+B \ct{bur03} system  the same four
post-Keplerian parameters as for PSR B1534+12 are available plus
$\dot P_b$ and a further constraint on $m_p/m_c$ coming from the
measurement of both the projected semimajor axes. On the contrary,
in the pulsar-white dwarf binary systems, which are the majority
of the binary systems with one pulsar and present almost circular
orbits, it is often impossible to measure $\dot\omega$ and
$\gamma$. Up to now, only $r$ and $s$ have been measured, with a
certain accuracy, in the PSR B1855+09 system \ct{kas94}, so that
it is impossible to use its data for testing the GTR as previously
outlined.
\subsection{The secular decrease of the mean anomaly and the binary pulsars}
%
%The utility of having at disposal a further post-Keplerian
%parameter seems apparent, in particular for the pulsar-white dwarf
%systems. Furthermore,
Let us investigate the magnitude of the mean anomaly precession in
some systems including one or two pulsars.

For PSR B1913+16 we have \cite{weistay05} $m_p=1.4414{\rm
M}_{\odot}$, $m_c=1.3867{\rm M}_{\odot},e=0.6171338,$
$P_b=0.322997448930$ d. Then, $\nu=0.2499064$, $F=1.04459537192$
and $\left.\dot\xi\right|_{\rm GE }=-10.422159$ deg yr$^{-1}$.
 For PSR
J0737-3039 A we have \cite{Kraetal05} $m_p=1.337{\rm M_{\odot}}$,
$m_c=1.250{\rm M_{\odot}}$, $e=0.087779$, $P_b=0.102251561$ d, so
that $\nu=0.249721953643$, $F=0.946329857430$. Thus,
$\left.\dot\xi\right|_{\rm GE }=-47.79$ deg yr$^{-1}$. This
implies that the ratio of the post-Newtonian gravitoelectric
secular rate of the mean anomaly to the mean motion amounts to
$\sim 10^{-5}$. Let us see if such post-Newtonian shift is
detectable from quadratic fits of the orbital phases of the form
$\mathcal{M}=a_0+b_0t+c_0t^2$. For PSR B1913+16 the quadratic
advance due to the gravitational wave emission over thirty years
amounts to ($\dot P_b=-2.4184\times 10^{-12} $)\eqi \Delta {
\mathcal{M} }=-\pi\left(\rp{\left.\dot P_b\right|_{\rm gw
}}{P^2_b}\right)(t-T_0)^2=0.5\ {\rm deg},\eqf with an uncertainty
$\delta(\Delta { \mathcal{M} } )$ fixed to $0.0002$ deg by
$\delta\dot P_b=0.0009\times 10^{-12}$. The linear shift due to
\rfr{fine}  amounts to
\eqi\Delta{\mathcal{M}}=\left.\dot\xi\right|_{\rm GE
}(t-T_0)=-312.6647\ {\rm deg}\eqf over the same time interval. The
uncertainty in\footnote{The sum of the masses and the semimajor axis entering $n$ are determined from timing data processing independently of $\mathcal{M}$ itself, e.g. from the periastron rate and the projected barycentric semimajor axis.} $n$ amounts to $1\times 10^{-9}$ deg yr$^{-1}$ due
to $\delta P_b=4\times 10^{-12}$ d. For PSR J0737-3039 A
 the gravitational wave emission over three years
($\dot P_b=-1.20\times 10^{-12}$) induces a quadratic shift of
$0.008$ deg, with an uncertainty $\delta(\Delta { \mathcal{M} } )$
fixed to $0.0005$ deg by $\delta\dot P_b=0.08\times 10^{-12}$. The
linear shift due to \rfr{fine} amounts to -143.3700 deg over the
same time interval. The uncertainty in $n$ amounts to $7\times
10^{-7}$ deg yr$^{-1}$ due to $\delta P_b=2\times 10^{-10}$ d.
Thus,  it should be possible to extract
$\left.\dot\xi\right|_{\rm GE}$ from the measured coefficient
$b_0$; both the corrupting bias due to the uncertainties in the
quadratic signature and the errors in $n$ would be negligible.

Measuring $\left.\dot\xi\right|_{\rm GE}$ as a further
post-Keplerian parameter would be very useful in those scenarios
in which some of the traditional post-Keplerian parameters are
known with a modest precision or, for some reasons, cannot be
considered entirely reliable\footnote{The measured value of the
derivative of the orbital period $\dot P_b$ is aliased by several
external contributions which often limit the precision of the
tests of competing theories of gravity based on this
post-Keplerian parameter \ct{sta03}. }. E.g., in the double pulsar
system PSR J0737-3039 A+B the parameters $r$ and $\gamma$ are
measured with a relatively low accuracy \ct{lyn04}. Moreover,
there are also pulsar binary systems in which only the periastron
rate has been measured \ct{kas99}: in this case the knowledge of
another post-Keplerian parameter would allow to determine the
masses of the system, although it would not be possible to
constraint alternative theories of gravity.

\section{Conclusions}
In this paper we have analytically derived for a
two-body system in eccentric orbits the secular variation
$\left.\dot\xi\right|_{\rm GE}$ yielding the  post-Newtonian
general relativistic gravitoelectric part of the precession of the mean anomaly not due to the variation of the orbital period.
In the limit of small
eccentricities and taking the mass of one of the two bodies
negligible, our results have been compared to the outcome of a
numerical integration of the post-Newtonian general relativistic
gravitoelectric equations of motion of the planets of the Solar
System performed by JPL: the agreement between the analytical and
numerical calculation is complete. Subsequently, we have
investigated the possibility of applying the obtained results to
the binary systems in which one pulsar is present. In particular,
it has been shown that the variation of the orbital period
$\left.\dot P_b\right|_{\rm gw}$ due to gravitational wave
emission and the effect derived by us are different ones. Indeed,
the post-Newtonian gravitoelectric precession of the mean anomaly,
which is always negative, is related to the secular increase of
the time of pericentre passage and occurs even if the orbital
period does not change in time. A quadratic fit of the orbital
phase of the pulsar would allow to measure
$\left.\dot\xi\right|_{\rm GE}$ because the biases due to the
errors in the quadratic shift due to $\dot P_b$ and in the linear
shift due of the mean motion $n$ are smaller. The use of
$\left.\dot\xi\right|_{\rm GE}$ as a further post-Keplerian
parameter would allow to improve and enhance the tests of
post-Newtonian gravity especially for those systems in which only
few post-Keplerian parameters can be reliably measured.

\section*{Acknowledgments}
I am grateful to E.M. Standish (JPL) for the results of his
numerical calculations about the post-Newtonian secular rates of
the planetary mean anomalies.  I also thank the anonymous referee
for his/her useful comments which improved the manuscript.

%-----------------------------------------

%-------------------------------------------------
\end{document}